\documentclass[aps,prb,twocolumn,showpacs,amsmath,amssymb,superscriptaddress,longbibliography,floatfix]{revtex4-2}
\usepackage{blindtext}
\usepackage{amsmath}
\usepackage{amssymb}
\usepackage{graphicx}
\usepackage{comment}
\usepackage{commath}
\usepackage{physics}
\usepackage{url}
\usepackage{xcolor}
\usepackage{multirow}
\usepackage[breaklinks,colorlinks=true,citecolor=blue]{hyperref}
\usepackage{xcolor}

\begin{document}
\title{Current cross-correlation spectroscopy of Majorana bound states}

\author{Michael Ridley}
\email{mikeridleyphysics@gmail.com}
\affiliation{Faculty of Engineering and the Institute of Nanotechnology and Advanced Materials,
Bar-Ilan University, Ramat Gan 5290002, Israel}

\author{Eliahu Cohen}
\affiliation{Faculty of Engineering and the Institute of Nanotechnology and Advanced Materials,
Bar-Ilan University, Ramat Gan 5290002, Israel}

\author{Christian Flindt}
\affiliation{Department of Applied Physics, Aalto University, 00076 Aalto, Finland}

\author{Riku Tuovinen}
\affiliation{Department of Physics, Nanoscience Center, University of Jyv{\"a}skyl{\"a}, 40014 Jyväskylä, Finland}

\begin{abstract}
The clock speed of topological quantum computers based on Majorana zero mode (MZM)-supporting nanoscale devices is limited by the time taken for electrons to traverse the device. We employ the time-dependent Landauer-B{\"u}ttiker transport theory for current cross-lead correlations in a superconducting nanowire junction hosting MZMs. From the time-dependent quantum noise, we are able to extract traversal times for electrons crossing the system. After demonstrating a linear scaling of traversal times with nanowire length, we present a heuristic formula for the traversal times which accurately captures their behaviour. We then connect our framework to a proposed experimental verification of this discriminant between spurious and genuine MZMs utilizing time-resolved transport measurements.
\end{abstract}

\maketitle

\section{Introduction}

Modern topological computing protocols are increasingly dominated by the exchange and manipulation of non-local, zero-energy, zero-charge, zero-spin quasiparticles known as Majorana Zero Modes (MZMs) \cite{kitaev2003fault,clarke2008superconducting,alicea2010majorana}. Recent experimental work has demonstrated the ability to engineer MZMs in superconducting heterostructures, particularly those based on
InAs or InSb \cite{mourik2012signatures, krogstrup_epitaxy_2015, lutchyn_majorana_2018, flensberg2021engineered}, and more recently Ge \cite{scappucci2021germanium}. 

\begin{figure}[t]
\includegraphics[width=0.95\columnwidth]{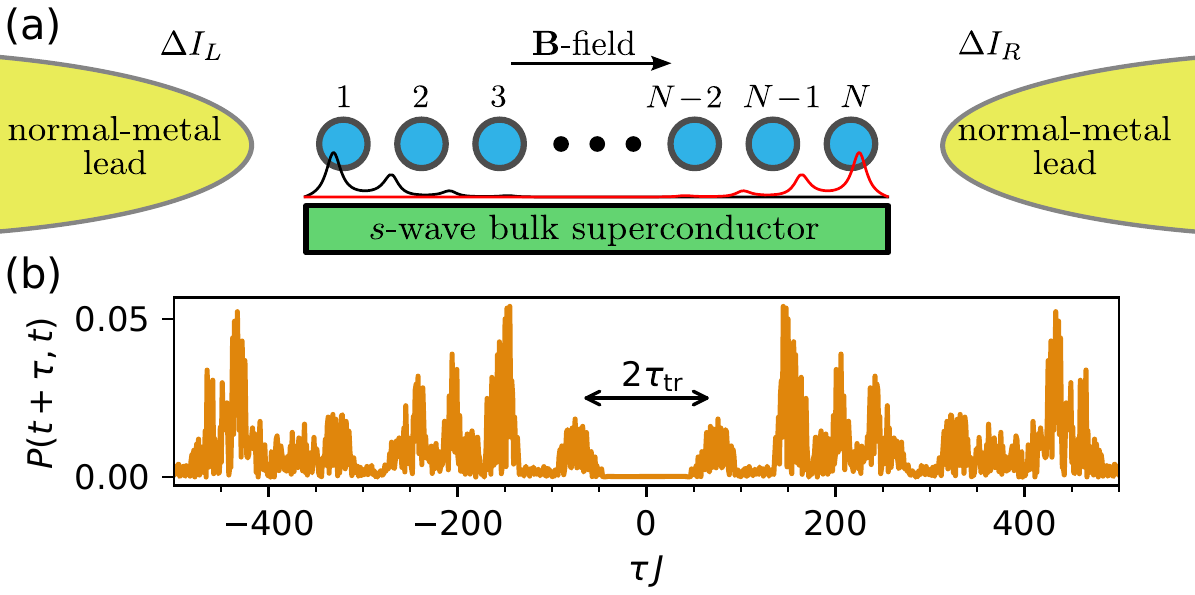}
\caption{(a) The superconducting nanowire in the presence of an external magnetic field and coupled to normal metal leads, $L$ and $R$. The black and red lines denote schematically the exponentially localized probability density of the Majorana zero mode. (b) Schematic procedure of extracting the traversal time from the distance of current cross-correlation peaks. 
}
\label{fig:nanowire}
\end{figure}

Among the most promising systems for realizing MZMs in the laboratory are proximitized nanowires~\cite{oreg2010helical, mourik2012signatures, rokhinson2012fractional, das2012zero, deng2012anomalous, wu2012tunneling, churchill2013superconductor, anh2024large, badawy2024electronic}, an example of which is shown schematically in Fig.~\ref{fig:nanowire}(a). The
MZMs’ non-Abelian anyonic statistics and their exponential localization at the opposite ends of the nanowire \cite{albrecht2016exponential} are highly desired
properties for designing quantum computation with reduced decoherence issues due to topological protection. The main signature of MZMs in normal metal--superconductor--normal metal junctions is a conductance peak at zero energy~\cite{finck2013anomalous, suominen2017zero}. However, the presence of disorder or Andreev bound states in
semiconductor--superconductor hybrid devices can cause spurious zero energy modes which are difficult to distinguish from the genuine quasiparticle~\cite{tripathi2016fingerprints, stanescu2018building, woods2021charge, sarma2023, ghosh2024majorana, laubscher2024majorana}, disrupting the conclusive observation of MZMs thus far. As such, experimental techniques with the sensitivity to make this distinction are sorely needed~\cite{stanescu2013majorana}.

In this work, we use dynamical current correlations and electron traversal times~\cite{ridley2019electron,ridley2022many} to identify Majorana zero modes (see Fig.~\ref{fig:nanowire}(b)). Earlier studies have explored noise in MZM-supporting nanowires~\cite{bolech2007observing, zocher2013modulation, beenakker2020shot, manousakis2020weak,manna2024full,mondal2025distinguishing}, and high-frequency noise and cross-correlation measurements enable time-resolved detection in mesoscopic conductors~\cite{Schoelkopf:1997, Reulet:2003, Gabelli:2008, Xue:2009, ubbelohde2012measurement, Ranni:2021, Brange:2021, Bayer:2025}. Moreover, shot noise and correlation measurements have been proposed as a basis for discriminating genuine MZMs in recent publications \cite{manousakis2020weak,cao2023differential}. Motivated by the noninvasive Fano-factor approach~\cite{manousakis2020weak}, we show that the nonequilibrium current cross-correlation yields a parameter-tolerant signature present for topological MZMs and absent for trivial states. To our knowledge, these transient Majorana signatures have not been identified previously.

The nonequilibrium Green's function methodology \cite{stefanucci2013nonequilibrium} is well-suited to the study of electron transport in superconducting nanowires \cite{stefanucci2010time, wu2012tunneling, tuovinen2019distinguishing, sriram2019supercurrent, ridley2022many, mondal2025transport}.
Recent years have seen the development of various approaches to time-dependent quantum transport based on the Green's function methodology, including the time-dependent Meir-Wingreen~\cite{meir_landauer_1992, jauho_time-dependent_1994, myohanen_kadanoff-baym_2009} and Landauer-B{\"u}ttiker (TDLB) approaches~\cite{buttiker_generalized_1985, tuovinen2016time, ridley2017time}, as well as the time-linear generalized Kadanoff-Baym ansatz for modeling correlated dynamics~\cite{tuovinen2023time, pavlyukh2024cheers, tuovinen2024electroluminescence, pavlyukh2025open, tuovinen2025thermoelectric}. Notably, the TDLB theory is a single-shot approach where time steps can be evaluated independently of each other, which enables trivial parallelization, making it computationally more attractive even compared to the time-linear theory.

The TDLB approach models transient correlations following a partition-free quench process~\cite{ridley2018formal}, where the leads equilibrate with the nanowire prior to the addition of a bias across the leads at the switch-on time $t_{0}$, and can deal with arbitrary time-dependent biases~\cite{ridley2016fluctuating}. It includes the study of time-dependent quantum noise on the signal via calculation of the cross-lead current correlation function~\cite{ridley2017partition,ridley2019electron} 
\begin{equation}\label{eq:CLR}
C_{LR}\left(t+\tau,t\right)\equiv \left\langle \Delta\hat{I}_{L}\left(t+\tau\right) \Delta\hat{I}_{R}\left(t\right)\right\rangle,
\end{equation}
where $\Delta\hat{I}_{\gamma}\left(t\right)\equiv\hat{I}_{\gamma}\left(t\right)-\left\langle\hat{I}_{\gamma}\left(t\right)\right\rangle$ is the fluctuation of the electronic current in lead $\gamma\in\{L,R\}$. In this expression, $t$ is the \emph{observation time} and $\tau$ is the \emph{relative time}. 
The experimentally-accessible cross correlation between the left and right leads is obtained as the real, lead-symmetrized average
\begin{equation}\label{eq:p}
P(t+\tau,t) \equiv \frac{1}{2}\,\mathrm{Re}\!\left[C_{LR}(t+\tau,t)+C_{RL}(t+\tau,t)\right].
\end{equation}
It is shown explicitly in Appendix \ref{sec:Appendix_A} that
\begin{equation}\label{eq:p_symmetry}
\mathrm{Re}[C_{LR}(t+\tau,t)]=\frac{1}{2}\langle\{\Delta\hat I_L(t+\tau),\Delta\hat I_R(t)\}\rangle
\end{equation}
and therefore $P$ coincides with the Blanter-B\"uttiker cross correlation in reciprocal, stationary settings ($\mathrm{Re}\,C_{LR}=\mathrm{Re}\,C_{RL}$)~\cite{blanter_shot_2000}. Also in non-reciprocal cases, Eq.~\eqref{eq:p} provides a lead-exchange symmetric measure of cross correlation. Under stationarity, all quantities depend only on $\tau$, and the associated Fourier transformation, $P(\omega)$, is typically referred to as noise power.

While a dc Fano factor based protocols can provide valuable diagnostics in appropriately designed setups \cite{manousakis2020weak,cao2023differential}, they are often time-averaged and do not directly resolve propagation delays. Here we therefore focus on the full time-domain cross-lead correlator, whose echo structure separates prompt local processes from delayed nonlocal ones. It also enables an operational extraction of the traversal time across the wire --- a genuinely nonlocal signature expected when spatially separated Majoranas dominate transport \cite{bolech2007observing,nilsson2008splitting}. Consistently, finite-frequency or waiting-time diagnostics predict qualitative differences between topological and trivial cases, reinforcing the advantage of the time domain as a discriminator~\cite{mi2018electron, bathellier2019finite, smirnov2019majorana}.

In the following, we develop the two-time cross-lead current-current correlator in the TDLB-Nambu framework and show that its time-domain echoes yield a length-linear traversal time containing a nonlocal delay robust to trivial near-zero modes. This builds on Ref.~\cite{tuovinen2019distinguishing}, where time-resolved current signatures were studied. The extra contributions to the traversal times are well captured within a novel and simple-to-implement heuristic formula. Finally, we propose an experiment where this intuitive approach can be used to characterize genuine MZMs.

\section{The model}

The non-interacting TDLB picture can be extended to describe a proximitized superconducting nanowire, where superconductivity is induced by a bulk $s$-wave superconductor as shown schematically in Fig.~\ref{fig:nanowire}(a). This is a $1D$ nanowire sandwiched between two normal metal conductors, featuring a strong spin-orbit interaction characteristic of, e.g. InSb or Ge. The spin therefore naturally aligns with the $\pm y$-direction, however, a strong external applied $B$-field along the axis of the nanowire breaks time-reversal symmetry and cants the spins along the $\pm z$-direction, introducing a Zeeman splitting $V_{Z} = g\mu_{B}B/2$, where $\mu_{B}$ is the Bohr magneton and $g$ is the Land{\'e} factor.

In Bardeen-Cooper-Schrieffer theory, the phonon-mediated attraction between electrons is captured by the pairing field $\Delta$~\cite{bardeen1957theory}. This translates to a tractable single-particle picture using a Nambu-spinor representation. We let operators $\hat{d}_{n\sigma}$ and $\hat{d}^{\dagger}_{n\sigma}$ annihilate and create electrons with spin $\sigma$, in spatial location $n$, and introduce a Nambu representation, $\hat{\Phi}_n = (\hat{d}_{n\uparrow},\hat{d}_{n\downarrow}^\dagger,\hat{d}_{n\downarrow},\hat{d}_{n\uparrow}^\dagger)^T$. The Hamiltonian of the isolated wire is given in the Bogoliubov–de Gennes form \cite{zhu2016bogoliubov}
\begin{equation}\label{eq:Wire_Hamiltonian}
    \hat{H}_{w}=\frac{1}{2}\underset{n}{\sum}\left[\hat{\Phi}_{n}^{\dagger}A_{n}\hat{\Phi}_{n}+\left(\hat{\Phi}_{n}^{\dagger}B_{n}\hat{\Phi}_{n}+\text{h.c.}\right)\right],  
\end{equation}
where the on-site contribution is given by
\begin{equation}
A_{n}=\left(\begin{smallmatrix}
J-\mu+V_{Z} & -\Delta & 0 & 0\\
-\Delta & \mu-J+V_{Z} & 0 & 0\\
0 & 0 & J-\mu+V_{Z} & \Delta\\
0 & 0 & \Delta & \mu-J+V_{Z}
\end{smallmatrix}\right)
\end{equation}
and the intra-wire hopping contribution is the matrix
\begin{equation}
B_{n}=\left(\begin{smallmatrix}
-J/2 & 0 & -\alpha/2 & 0\\
0 & J/2 & 0 & -\alpha/2\\
\alpha/2 & 0 & -J/2 & 0\\
0 & \alpha/2 & 0 & J/2
\end{smallmatrix}\right),
\end{equation}
with parameters $J$, $\mu$ and $\alpha$ describing the hopping, chemical potential and spin-orbit coupling strength of the nanowire, respectively. For reference, material parameters typical of proximitized InSb nanowires lie in the ranges $J \simeq 10$--$100~\mathrm{meV}$, $\mu \simeq 0$--$5~\mathrm{meV}$, $\alpha \simeq 0.5$--$1~\mathrm{eV{\cdot}\mathring{A}}$, $V_Z \simeq 0.2$--$3~\mathrm{meV}$, and $\Delta \simeq 0.2$--$0.3~\mathrm{meV}$~\cite{mourik2012signatures, krogstrup_epitaxy_2015, lutchyn_majorana_2018}.

The energy levels of the leads are subjected to a time-dependent, spatially homogenous shift by a voltage at time $t_{0}$. That is, $\varepsilon_{kL/R}\rightarrow\varepsilon_{kL/R}+V_{L/R}\left(t\right)$ at $t>t_{0}$. The Hamiltonian of the leads is then expanded in the Nambu
basis as $\hat{H}_{l}=\frac{1}{2}\underset{k\gamma}{\sum}\hat{\Phi}_{k\gamma}^{\dagger}\tilde{\varepsilon}_{k\gamma}\hat{\Phi}_{k\gamma}$, where $\tilde{\varepsilon}_{k\gamma}\equiv\varepsilon_{k\gamma}\textrm{diag}\left(1,-1,1,-1\right)$ is the energy of state $k$ in the $\gamma$-th lead. 

Finally, the lead-nanowire coupling Hamiltonian is given by $\hat{H}_{c}=\frac{1}{2}\underset{nk\gamma}{\sum}\left(\hat{\Phi}_{n}^{\dagger}\tilde{T}_{nk\gamma}\hat{\Phi}_{k\gamma}+h.c\right)$, where $\tilde{T}_{nk\gamma}\equiv T_{nk\gamma}\textrm{diag}\left(1,-1,1,-1\right)$ are the tunneling matrix elements between the $k$-th state of lead $\gamma$ and the $n$-th state of the wire. The Hamiltonian of the total system then reads $\hat{H}_{\mathrm{tot}}=\hat{H}_w+\hat{H}_l+\hat{H}_c$.

It is worth pointing out that in the present model the pairing field $\Delta$ in $\hat H_w$ is an effective, static parameter describing proximity-induced superconductivity. The underlying bulk $s$-wave superconductor is therefore not included as an explicit transport terminal. Consequently, the only reservoirs that drive the dynamics and define the current operators are the two normal leads $L$ and $R$, and throughout we focus on their currents and the corresponding cross correlations. We also neglect charging-energy physics and interactions more generally~\cite{fu_electron_2010, tuovinen2021electron, souto_majorana_2025}. Within our model, the nonzero cross correlations originate from coherent nonlocal scattering processes that connect the two normal contacts through the proximitized wire.

\section{The method}\label{sec:method}

We generalize the TDLB approach to the quantum noise~\cite{ridley2017partition} so that all Green's functions appearing in the current-current correlation function are given in the Nambu representation, where each of the $N$-sites on the nanowire shown in Fig.~\ref{fig:nanowire}(a) corresponds to four Nambu states. For inter-site spacing $a$ the wire length $L=Na$. The one-particle Green's function in the Nambu representation is defined as 
\begin{equation}
\left[\mathbf{G}\left(z_{1},z_{2}\right)\right]_{mn}=-i\left\langle \mathcal{T}_{KP}\left[\hat{\Phi}_{m}\left(z_{1}\right)\otimes\hat{\Phi}_{n}^{\dagger}\left(z_{2}\right)\right]\right\rangle,
\end{equation}
where $\mathcal{T}_{KP}$ denotes an ordering of times $z_{1}$, $z_{2}$ along the complex Konstantinov-Perel' contour \cite{konstantinov1960graphical}, and the brackets denote a quantum statistical average. In the noninteracting case, this contour is equivalent to the standard Keldysh contour \cite{ridley2018formal}; however the vertical branch on the Konstantinov-Perel' contour allows for a more intuitive treatment of coupled subsystems prior to the quench time $t_{0}$. We set $\hbar=e=k_{\mathrm{B}}=1$ throughout this work. Quantities denoted in bold are $4N\times 4N$-dimensional matrices. $\mathbf{G}$ satisfies the Kadanoff-Baym integro-differential equations of motion on the contour \cite{baym1961conservation}, with integral kernel given by the embedding self-energy $\left[\mathbf{\Sigma}_{em}\left(z_{1},z_{2}\right)\right]_{mn}=\underset{k \gamma}{\sum}\tilde{T}_{mk\gamma}\left[\mathbf{g}_{\gamma\gamma}\left(z_{1},z_{2}\right)\right]_{kk}\tilde{T}_{k\gamma n}$, where $\mathbf{g}_{\gamma\gamma}$ denotes the `bare' Green's function of lead $\gamma$ and the $\tilde{T}_{mk\gamma}$ are defined in the lead-nanowire coupling $\hat{H}_{c}$. All self-energy components can be expressed in terms of $\mathbf{\Gamma}_{\gamma}=2\pi\sum_{k}\tilde{T}_{m k\gamma}\tilde{T}_{k\gamma n}\delta\left(\varepsilon_{\gamma}^{F}-\tilde{\varepsilon}_{k\gamma}\right)$, where $\varepsilon_{\gamma}^{F}$ is the Fermi energy of lead $\gamma$ \cite{ridley2015current}. The level-width $\mathbf{\Gamma}=\sum_\gamma \mathbf{\Gamma}_\gamma$ describes the energy-independent lead-nanowire coupling within the wide-band limit approximation~\cite{covito2018transient}. This approximation is valid because we aim to study low-energy excitations near the Fermi level and the normal metal leads can be safely assumed to have broad energy bands compared to the Majorana energy scale.

The lesser and greater components of the one-particle Green's function of the nanowire are given by 
\begin{equation}\label{eq:GF_lessgtr}
    \mathbf{G}^{\lessgtr}\left( t_{1},t_{2} \right)=\pm i\int\frac{d\omega}{2\pi}f\left(\pm\omega\right)\underset{\gamma}{\sum}\mathbf{S}_{\gamma}\left(t_{1};\omega\right)\mathbf{\Gamma}_{\gamma}\mathbf{S}_{\gamma}^{\dagger}\left(t_{2};\omega\right),
\end{equation}
where $f\left(x\right)=\left(e^{\beta \left(x-\mu\right)}+1\right)^{-1}$ is the Fermi-Dirac distribution, $\beta$ is the inverse temperature and $\mu$ is the chemical potential \cite{ridley2015current}.

We have introduced the matrix $\mathbf{S}_{\gamma}\left(t;\omega\right)\equiv e^{-i\mathbf{h}^{\text{eff}}\left(t-t_{0}\right)}\left[\mathbf{G}^{r}\left(\omega\right) -i\int_{t_{0}}^{t}d\bar{t}e^{-i\left(\omega\mathbf{I}-\mathbf{h}^{\text{eff}}\right)\left(\bar{t}-t_{0}\right)}e^{-i\psi_{\gamma}\left(\bar{t},t_{0}\right)}\right]$ in terms of the retarded Green's function $\mathbf{G}^{r}\left(\omega\right)=\left(\omega\mathbf{I}-\mathbf{h}^{\text{eff}}\right)^{-1}$, where $\mathbf{h}^{\text{eff}}=\mathbf{h}-i\mathbf{\Gamma}/2$ is the effective Hamiltonian of the nanowire considered as an open quantum system, reducing to the `bare' nanowire Hamiltonian $\mathbf{h}$, i.e. expressing $\hat{H}_w$ in the Nambu basis, in the decoupling limit $\mathbf{\Gamma}\rightarrow0$. 
We also introduced the bias-voltage phase factor $\psi_{\gamma}\left(t_{1},t_{2}\right)\equiv{\int}_{t_2}^{t_1}d\tau\,V_{\gamma}\left(\tau\right)$. The two-time current correlator may be evaluated as a closed function in the two-time plane:
\begin{align}
& \label{eq:Correlation}
C_{LR}\left(t_{1},t_{2}\right) \nonumber \\
& =	4\mathrm{Tr}\left\{ \mathbf{\Gamma}_{L}\mathbf{G}^{>}\left(t_{1},t_{2}\right)\mathbf{\Gamma}_{R}\mathbf{G}^{<}\left(t_{2},t_{1}\right)\right. \nonumber \\
& \left.+ i\mathbf{G}^{>}\left(t_{1},t_{2}\right)\left[\mathbf{\Lambda}_{R}^{+}\left(t_{2},t_{1}\right)\mathbf{\Gamma}_{L}+\mathbf{\Gamma}_{R}\left(\mathbf{\Lambda}_{L}^{+}\right)^{\dagger}\left(t_{1},t_{2}\right)\right]\right.\nonumber \\
& \left.+i\left[\mathbf{\Lambda}_{L}^{-}\left(t_{1},t_{2}\right)\mathbf{\Gamma}_{R}+\mathbf{\Gamma}_{L}\left(\mathbf{\Lambda}_{R}^{-}\right)^{\dagger}\left(t_{2},t_{1}\right)\right]\mathbf{G}^{<}\left(t_{2},t_{1}\right) \right. \nonumber \\
& \left.-\mathbf{\Lambda}_{R}^{+}\left(t_{2},t_{1}\right)\mathbf{\Lambda}_{L}^{-}\left(t_{1},t_{2}\right)-\left(\mathbf{\Lambda}_{L}^{+}\right)^{\dagger}\left(t_{1},t_{2}\right)\left(\mathbf{\Lambda}_{R}^{-}\right)^{\dagger}\left(t_{2},t_{1}\right)\right\}.
\end{align}
The trace in Eq.~\eqref{eq:Correlation} is taken with respect to the nanowire degrees of freedom. We introduce the lead matrices $\mathbf{\Lambda}^{+}$ and $\mathbf{\Lambda}^{-}$, which correspond to electrons and positively charged holes, respectively, propagating from the leads, as $\mathbf{\Lambda}_{\gamma}^{\pm}\left(t_{1},t_{2}\right) =	\pm i e^{-i\psi_{\gamma}\left(t_{1},t_{0}\right)}\int\frac{d\omega}{2\pi}f\left(\pm\omega\right) e^{-i\omega\left(t_{1}-t_{0}\right)}\mathbf{\Gamma}_{\gamma}\mathbf{S}_{\gamma}^{\dagger}\left(t_{2};\omega\right)$~\cite{ridley2017partition}. Physically, the terms in Eq.~\eqref{eq:Correlation} are given by lead-lead, lead-nanowire and nanowire-nanowire $e$-$h$ interference effects. Refs.~\cite{ridley2015current} and~\cite{ridley2017partition} demonstrate explicitly how the standard Landauer-B{\"u}ttiker formalisms for the current and current fluctuations, respectively, are recovered from the TDLB formalism in the long-time and static bias limits.

\section{Traversal times}

In the case of the two-terminal system shown in Fig.~\ref{fig:nanowire}(a), the electronic traversal time $\tau_{\mathrm{tr}}$ can be extracted directly from the lead-symmetrized correlator data, shown in Fig.~\ref{fig:nanowire}(b), as half of the distance separating the first resonant peaks along the $\tau$-axis. This can be defined more explicitly as the value of the relative delay $\tau$ at which the stationary current cross correlation of Eq.~\eqref{eq:p} reaches its maximum within a restricted domain $\mathcal{T}$:
\begin{equation} \label{eq:traversaltime}
\tau_{\mathrm{tr}} = \underset{\tau\in\mathcal{T}}{\mathrm{arg}\,\mathrm{max}} |P(t+\tau,t)| \quad (t\to\infty).
\end{equation}
We take the maximum argument over the restricted domain $\mathcal{T}=[\tau_{\mathrm{min}},\tau_{\mathrm{max}}]$, chosen to exclude trivial near-zero features (before the clean ballistic transit) and later-time reflections discussed in Section \ref{sec:results}. For a simplified model nanowire, $\mu=\alpha=V_Z=\Delta=0$, $\tau_{\mathrm{tr}}(k)=L/v_g(k)$ with $L=Na$ and the group velocity $v_g(k)=d\omega_k/dk = Ja\sin(ka)$. The fastest mode occurs at $ka=\pi/2$ leading to $\tau_{\mathrm{min}}=N/J$. The upper bound may be set, e.g., to $\tau_{\mathrm{max}}=2\tau_{\mathrm{min}}$ to isolate the first through-wire peak.

We note that many closely related definitions for traversal times and similar quantum timescales exist in the literature~\cite{buttiker1982traversal, landauer1987diffusive, sokolovski1987traversal, nazarov1991influence, fertig1993path, landauer1994barrier, nitzan2000tunneling, peskin2002traversal, splettstoesser2009two, galapon2012only, pollak2017quantum, rivlin2021determination, he2022transition}. Our definition is intended to ground the traversal time in terms of a quantity that is already accessible to experimentalists, who can construct the cross correlation from repeated non-invasive transport measurements in the leads~\cite{jordan2005continuous, dicarlo2006system, gustavsson2006counting, gustavsson2009electron, ubbelohde2012measurement, von2020two}.

\section{Results}\label{sec:results}

We carry out calculations of the time-dependent cross-correlations for four distinct transport regimes, each specified by a choice of the parameters in our model of a nanowire. The parameters $\mu$, $J$, $V_{Z}$ and $\Delta$ defined above are sufficient to characterize the ordinary and topological superconducting regimes. In addition we study the case of magnetic impurities situated at the wire ends which is characterized by the modified tight-binding parameter $\tilde{V}_{Z}$ described in Ref. \cite{tuovinen2019distinguishing}. The spurious case of the quasi Majorana mode-supporting nanowire is defined by the usual parameters and an additional smooth confining potential within the nanowire (see Eq.~(13) in Ref.~\cite{tuovinen2019distinguishing}). All energies are measured with respect to $J$; times and temperatures are given in terms of $J^{-1}$.

\begin{table}[htbp]
\centering
\begin{tabular}{ |p{3cm}||p{0.8cm}|p{0.8cm}|p{0.8cm}|p{0.8cm}|p{0.8cm}|  }
 \hline
 \multicolumn{6}{|c|}{Parameter sets (in units of $J$)} \\
 \hline
 Model Regime & $\mu/J$ & $\alpha/J$ & $V_{Z}/J$ & $\tilde{V}_{Z}/J$ & $\Delta/J$ \\
 \hline
 Ordinary phase   & 0.0    & 0.5 &  0.0 & N/a & 0.1 \\
 Topological phase &  0.0  & 0.5 & 0.25 & N/a & 0.1 \\
 Magnetic impurity & 0.0 & 0.5 & 0.0 & 0.67 & 0.1\\
 Quasi Majorana    & 2.0 & 0.5 & 1.2 & N/a & 0.1\\
 \hline 
\end{tabular}
\caption{Parameter sets for the different model regimes considered in this work. 
}
\label{tab:parameter_sets}
\end{table}

The parameters chosen for our calculations are displayed explicitly in Table \ref{tab:parameter_sets}. In these situations, low-energy states arise from different physical mechanisms with distinct phenomenology. The distinct equilibrium energy spectra for the four regimes considered here are displayed for reference in Fig. \ref{fig:spectral} of Appendix \ref{sec:eq_spectral}. In the trivial superconducting phase, the spectrum is fully gapped and any subgap features correspond to conventional Andreev bound states, which are sensitive to local perturbations~\cite{laubscher_majorana_2021}. By contrast, MZMs in the topological phase appear as spatially separated, bound, exponentially-decaying end states pinned near zero energy, producing robust zero-bias conductance features and exhibiting nonlocal correlations~\cite{schiela_progress_2024}. Magnetic impurities can induce states occasionally crossing zero energy, but these remain strongly localized and tunable only via local magnetic parameters, lacking topological protection~\cite{schneider_precursors_2022}. Also, smooth confinement potentials can produce quasi-Majorana states that mimic many spectroscopic signatures of true MZMs, including apparent zero-bias robustness, yet differ in their strong dependence on confinement geometry and absence of nonlocal parity correlations~\cite{vuik_reproducing_2019}. Only the low-energy in-gap states are excited by applying a sudden bias voltage $V_L(t)=-V_R(t)=V_0\theta(t)$ with $V_0=\Delta/2$.

We then compute the cross-correlation $P\left(t+\tau,t\right)$ for each parameter set, and present its absolute value normalized by the square of the lead-nanowire coupling strength $\Gamma=0.01$ defined between the two end-sites of the wire and the lead, for each model regime in Fig.~\ref{fig:correlations}. Details of the numerical implementation can be found in Appendix B of Ref.~\cite{ridley2022many}. Crucially for our method, the implemented formula is a `single-shot' function in the two-time plane, so there is an identical cost associated with taking the observation time $t$ to be very small or very large. In practice, we set the observation time to $t=5000$, i.e. we work directly in the steady-state regime. This is demonstrated to be an observation time at which the cross-correlations have comfortably saturated in Appendix~\ref{sec:long_t}.

We also consider the low-temperature regime (to avoid thermal suppression of topological features) by setting $\beta=200$. With the hopping energies $J$ in sub-eV range for prototypical semiconductor nanowires~\cite{marramajorana2022}, this corresponds to cryogenic temperatures $\lesssim 10$~K. The temperature and tunneling-rate dependence of the cross-correlations are examined further in Appendix \ref{sec:temperature_dep}, confirming that variations in these parameters do not qualitatively alter the cross-correlation picture.

\begin{figure}[t]
\includegraphics[width=\linewidth]{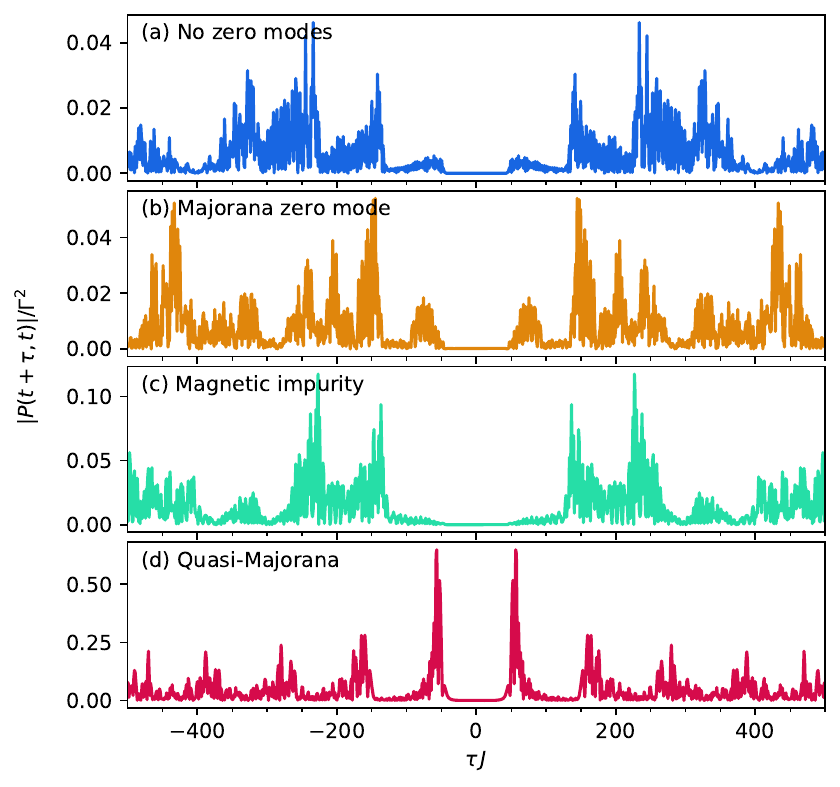}
\caption{Stationary current cross-correlation versus relative time shifts $\tau$ for superconducting nanowires of length $N=50$ sites. The nanowire is in the ordinary superconducting phase with no zero modes in panel (a) and it hosts in-gap states at zero energy for panels (b-d). Only the panel (b) hosts the topological Majorana zero mode while panels (c) and (d) host topologically trivial in-gap states. Coupling to the normal-metal leads is specified by the tunneling rate $\Gamma=0.01$, and the temperature is set by $\beta=200$.}
\label{fig:correlations}
\end{figure}

Immediately we see by comparing the case of the true topological superconductor (Fig.~\ref{fig:correlations}(b)) to the spurious Majorana case (Fig.~\ref{fig:correlations}(d)) that there is a qualitative distinction between the corresponding cross-correlation functions. Specifically, in Fig.~\ref{fig:correlations}(d) we observe strong resonant peaks at $\tau\simeq\pm 50$, which we identify as the coherent wavefront peaks, i.e., the earliest traversal signal across the nanowire. The zero value of the cross-correlation signal at times between these peaks corresponds to the fact that information transfer cannot occur between the metal leads on those timescales. These prominent early-time peaks are followed by an oscillatory peak structure at larger $|\tau|$, consistent with repeated passages through a finite system due to internal reflections. This is very similar to the behaviour observed in Ref.~\cite{ridley2019electron} for a graphene nanoribbon junction. In contrast, for the genuine topological superconductor in Fig.~\ref{fig:correlations}(b), the early-time wavefront peak is strongly suppressed. The dominant peak structure, with periodic repetition, builds up only on a significantly longer timescale. This general picture of the cross-correlation dynamics is confirmed in Appendix \ref{sec:length} for various nanowire lengths.

\begin{figure}[t]
  \includegraphics[width=\linewidth]{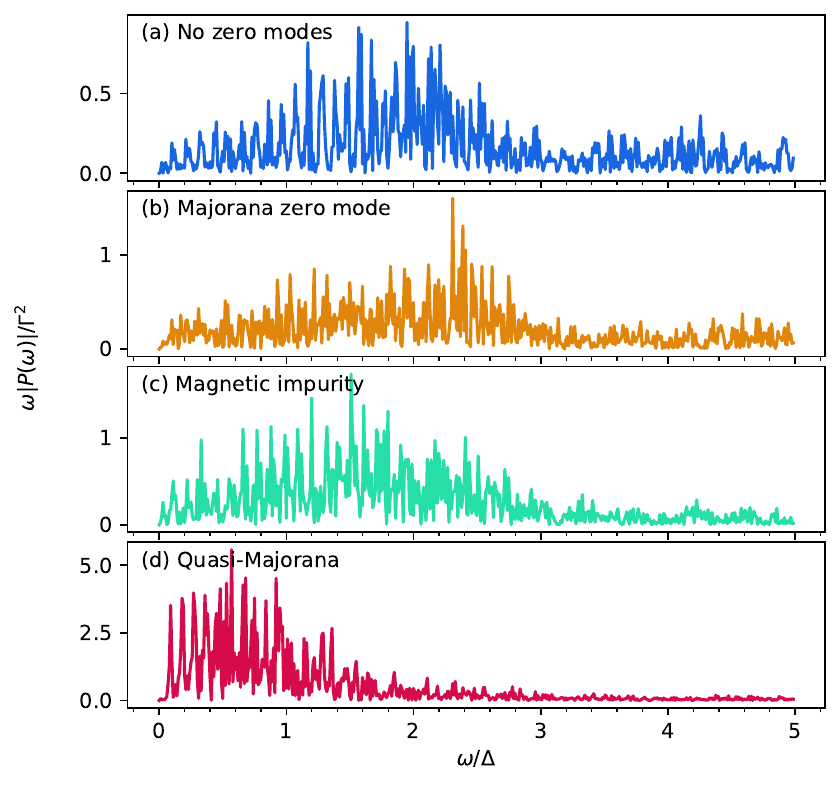}
  \caption{Frequency-resolved noise power spectra obtained from the time-dependent current cross correlations of Fig.~\ref{fig:correlations}.}
  \label{fig:fft}
\end{figure}

In relation to time-dependent current cross correlations of Eq.~\eqref{eq:p}, it is possible to perform a Fourier transform at the stationary state with respect to the relative time coordinate $\tau$,
\begin{equation}\label{eq:noisepower}
P(\omega)=\int_{-\infty}^\infty d \tau \, e^{i\omega\tau}P(t+\tau,t) \quad (t\to\infty),
\end{equation}
to obtain the noise power spectrum~\cite{blanter_shot_2000}. Figure~\ref{fig:fft} shows the weighted spectrum $\omega |P(\omega)|$ to suppress the generic low-frequency ($1/f$ type) contribution. We observe that the quasi-Majorana state hosts operational frequencies mostly within the superconducting
gap $\Delta$, which correspond to Andreev processes. In contrast, the MZM case shows a pronounced peak around $2\Delta$,
which would indicate coupling to continuum states (above the gap) or a pair process. The ordinary superconductor
and magnetic impurity cases instead show a broad range of operational frequencies.

\begin{figure}[t]
\includegraphics[width=\linewidth]{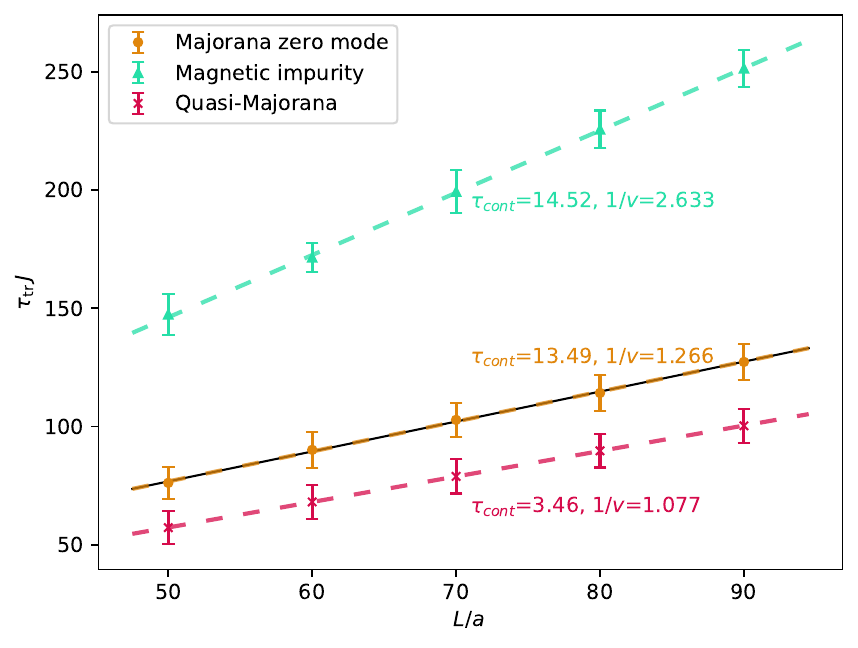}
  \caption{Extracted traversal times in terms of the nanowire length $L$. Simple linear fits (dashed lines) are performed as $\tau_{\mathrm{tr}}=\tau_{\mathrm{cont}}+L/v$, where $\tau_{\mathrm{cont}}$ and $1/v$ are obtained as the intercept and slope, respectively. For the Majorana zero mode, a fit with Eq.~\eqref{eq:heuristic} is also included (solid black line) with the parameters $\delta=1.402$ and $z=4$. Other model parameters are $\Gamma=0.01$ and $\beta=200$.}
  \label{fig:traversal-length}
\end{figure}

Figure~\ref{fig:traversal-length} shows the traversal times evaluated, in accordance with Eq.~\eqref{eq:traversaltime}, for each model regime as a function of the nanowire length $L$. This data clearly supports the interpretation of a traversal time, as the dependence on the wire length is linear. We can roughly decompose the traversal time as $\tau_{\mathrm{tr}}=\tau_0+\tau_{\mathrm{cont}}$, where $\tau_0$ captures the length-dependent wavefront contribution, while $\tau_{\mathrm{cont}}$ accounts for an additional delay associated with the end regions and contact-induced localization. In this interpretation, the suppression of the early-time wavefront peak in the topological regime (Fig.~\ref{fig:correlations}(b)) is reflected in a larger effective contact delay $\tau_{\mathrm{cont}}$, whereas in the quasi-Majorana regime the traversal signal is dominated by the length-dependent wavefront peak (Fig.~\ref{fig:correlations}(d)). The delay time $\tau_{\mathrm{cont}}$ is related to the localized edge states and how strongly they hybridize in the presence of leads, as discussed in Appendix \ref{sec:Appendix_heuristic}.

These observations can be translated into an heuristic formula for the traversal time if we evaluate it from the sum of the electron dwell times in a site-wise manner, i.e.
\begin{equation}
    \tau_{\mathrm{tr}}=\sum^{N}_{i=1}\tau_{i}.
\end{equation} 
A site-wise expression for the $\tau_{i}$ (given explicitly in Appendix \ref{sec:Appendix_heuristic}) is then evaluated by choosing an effective intra-site coupling $\lambda_{i}$ which follows the inverse of the on-site potential. The resulting heuristic expression is given by
\begin{equation}\label{eq:heuristic}
    \tau_{\mathrm{tr}}=\frac{4}{J\pi}\left[\frac{2\left(e^{\delta}-1\right)}{e^{\delta/z}-1}+N-2z\right],
\end{equation}
where $\delta$ controls the exponential drop-off at each end of the wire, $z$ is the localization length evaluated to the nearest integer, the on-site velocity is numerically estimated to be $v_{i}=\frac{\pi}{2} \lambda_{i}$ and $N=L/a$ is the wire length. Note that this expression scales linearly with $N$, as expected. 
The estimated traversal times $\tau_{\mathrm{tr}}\sim 100$ in units of inverse hopping would correspond to a temporal resolution of picoseconds, for prototypical semiconductor nanowires. Equation \eqref{eq:heuristic} offers a substantial computational speedup for the evaluation of traversal times in MZM-supporting nanowires compared to the more rigorous TDLB methodology presented in Section \ref{sec:method}.

\section{Conclusions}

In this work we analyzed time-resolved cross-lead current correlations in proximitized nanowire junctions within the TDLB framework, using a lead-symmetrized correlator to operationally define and discern an electronic traversal time from transient noise. We evaluated this diagnostic across conventional superconducting, topological Majorana, magnetic-impurity, and quasi-Majorana regimes, finding a robust linear growth of traversal time with wire length and regime-specific temporal fingerprints that distinguish genuine MZMs from spurious near-zero modes. We further distilled a compact heuristic capturing both length scaling and end-localization effects, and verified its quantitative accuracy against simulations. Finally, the predicted picosecond temporal scale expected for prototypical semiconductor nanowires indicates that traversal-time echoes should be observable with current experimental platforms.

To further estimate the time scales relevant for experimental observation, \citet{mourik2012signatures} reported an induced superconducting energy gap of about $250$~µeV in a hybrid superconductor-semiconductor nanowire device. This energy scale corresponds to characteristic in-gap oscillations with a period of roughly $16$~ps, or equivalently a frequency of about $60$~GHz. On the other hand, \citet{mciver_light-induced_2020} have demonstrated high-frequency transport measurements using photoconductive switches capable of sub-picosecond temporal resolution. These benchmarks suggest that the time scales highlighted by our calculations set a challenging but potentially accessible target for ultrafast transport experiments. However, we emphasize that realistic measurements will likely be substantially less clean than our idealized simulations, and that implementing time-resolved cross-correlations in a proximitized nanowire is experimentally demanding. We therefore view the time-domain echo structure identified here as a complementary, non-invasive handle on nonlocal transport, rather than a definitive standalone discriminator in all experimental circumstances. In practice, a possible route would be a stroboscopic two-lead implementation where time-resolved currents are detected at both contacts and the device is driven out of equilibrium in a controlled manner. While we focus on a voltage-bias quench in this work, the same framework can accommodate other driving protocols, including optical excitation.

Future studies could explore Floquet Majorana bound states~\cite{tong2013generating}. These can be implemented within the TDLB framework via the incorporation of a periodic driving voltage, offering an alternative way to manipulate and control topological states. Another interesting possibility is to apply the same time-resolved cross-lead-correlator diagnostic to platforms hosting \(\mathbb{Z}_n\) parafermionic zero modes~\cite{zamolodchikov1985nonlocal}. In such systems, fractionalized quasiparticles with effective charge \(e^\ast{=}2e/n\) and the associated \(4\pi/n\)-Josephson periodicity should imprint distinct traversal-time fingerprints, e.g., multiple delayed branches and fractional noise scaling, providing a clear discriminant from Majorana (\(n{=}2\)) responses and a route to time-domain identification of non-Abelian orders beyond MZMs.

\section*{Acknowledgments}

M.~R.~and E.~C.~acknowledge support from the European Union’s Horizon Europe research and innovation programme under grant agreement No.~101178170 and by the Israel Science Foundation under grant agreement No.~2208/24. C.~F.~acknowledges the financial support from the Research Council of Finland through the Finnish Centre of Excellence in Quantum Technology (Project No.~352925) and the Finnish Quantum Flagship (Project No.~358877). R.~T.~acknowledges the financial support of the Jane and Aatos Erkko Foundation (Project EffQSim) and the Research Council of Finland through the Finnish Quantum Flagship (Project No.~359240). We also acknowledge CSC---IT Center for Science, Finland, for computational resources.

\appendix
\begin{widetext}
\section{Two-time current correlation functions}\label{sec:Appendix_A}
The time-dependent Landauer-Büttiker approach~\cite{stefanucci2013nonequilibrium, tuovinen2014time, ridley2015current} involves the study of time-dependent quantum noise on the signal via calculation of the cross-lead current correlation function~\cite{ridley2017partition,ridley2019electron} 
\begin{equation}\label{eq:C_LR}
C_{LR}\left(t_1,t_2\right)\equiv \left\langle \Delta\hat{I}_{L}\left(t_1\right) \Delta\hat{I}_{R}\left(t_2\right)\right\rangle,
\end{equation}
where $\Delta\hat{I}_{\gamma}\left(t\right)\equiv\hat{I}_{\gamma}\left(t\right)-\left\langle\hat{I}_{\gamma}\left(t\right)\right\rangle$ is the fluctuation of the electronic current in lead $\gamma\in\{L,R\}$. In Eq.~\eqref{eq:C_LR}, $\Delta\hat{I}_L$ and $\Delta \hat{I}_R$ do not commute, in general, and the object is complex-valued following the symmetry property $C_{LR}^*(t_1,t_2)=C_{RL}(t_2,t_1)$~\cite{ridley2022many}. For this reason, it is advantageous to define the symmetrized function $P_{LR}(t_1,t_2)=\mathrm{Re}[C_{LR}(t_1,t_2)]$. Utilizing the complex number identity $2\mathrm{Re}[z]=z+z^*$, it can be shown that
\begin{equation}
P_{LR}(t_1,t_2)=\mathrm{Re}[C_{LR}(t_1,t_2)]=\frac{1}{2}\left\langle\left\{\Delta\hat{I}_L(t_1),\Delta\hat{I}_R(t_2)\right\}\right\rangle,
\end{equation}
i.e., it coincides with the symmetrized current correlation function of~\citet{blanter_shot_2000}. In the main text, we further introduce the experimentally-accessible lead-symmetrized cross-correlator as
\begin{equation}
P(t+\tau,t) = \frac{1}{2}\left[P_{LR}(t+\tau,t)+P_{RL}(t+\tau,t)\right] = \frac{1}{2}\mathrm{Re}\left[C_{LR}(t+\tau,t)+C_{RL}(t+\tau,t)\right],
\end{equation}
where $t$ is the observation time and $\tau\equiv t_1-t_2$ is the relative time. It is worth pointing out that performing a similar decomposition for the imaginary part,
\begin{equation}
\frac{1}{2}\mathrm{Im}\left[C_{LR}(t+\tau,t)+C_{RL}(t+\tau,t)\right]= \frac{1}{4i} \left[ C_{LR}(t+\tau,t)-C_{RL}(t,t+\tau) + C_{RL}(t+\tau,t)-C_{LR}(t,t+\tau) \right] ,
\end{equation}
does not lead to an object expressed directly in terms of measurable quantities.

\begin{figure}[h!]
  \includegraphics[width=0.6\linewidth]{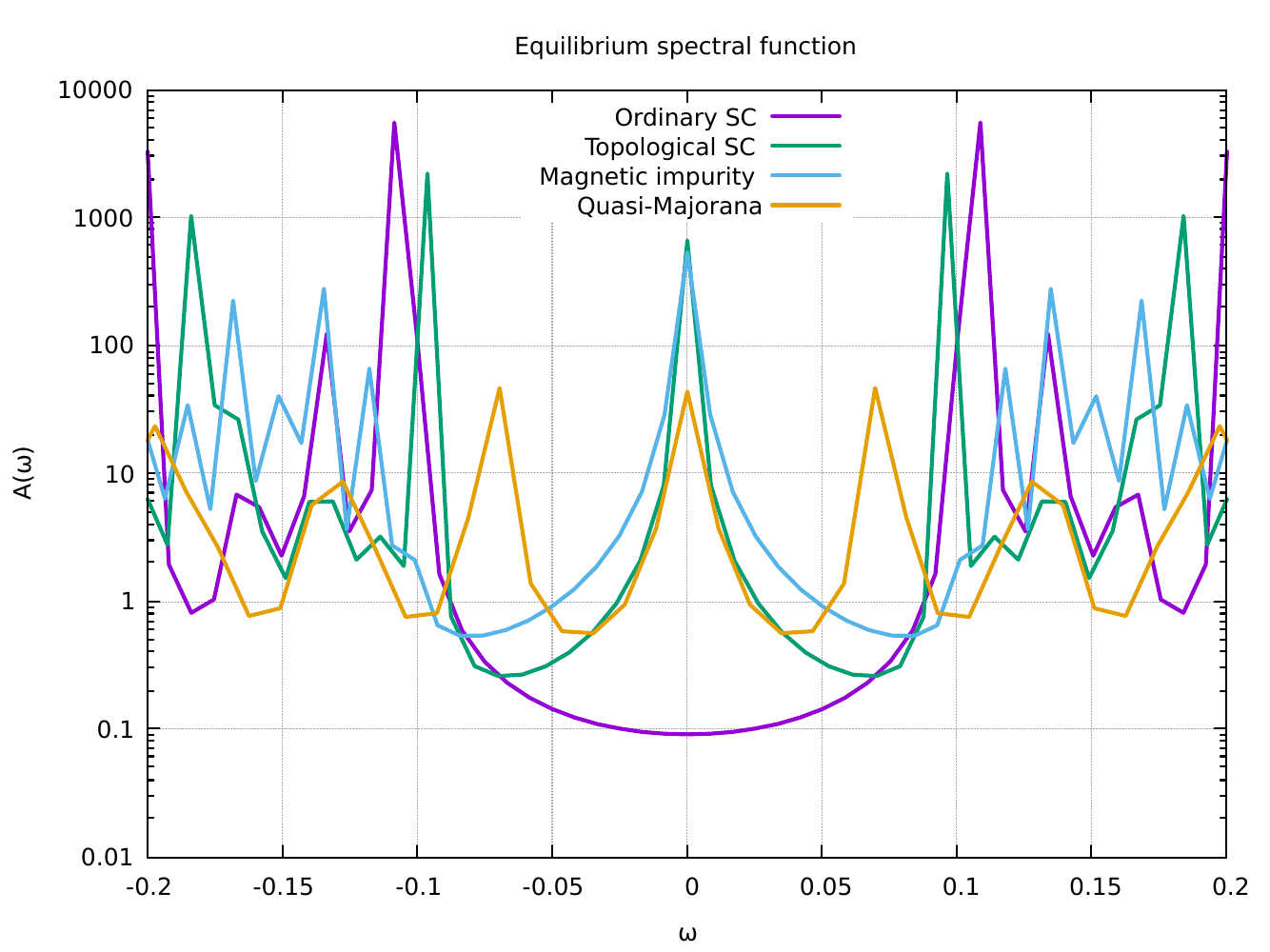}
  \caption{Equilibrium spectral function focused at the low-energy regime for the four cases studied in the main text. Notably, besides the ordinary superconductor, all other cases host in-gap states pinned to zero energy. The superconducting pair potential is $\Delta=0.1$.}
  \label{fig:spectral}
\end{figure}

\section{Equilibrium spectral function}\label{sec:eq_spectral}
In order to check that the superconducting nanowire hosts in-gap states, we consider the equilibrium spectral function
\begin{equation}
A(\omega) = -\frac{1}{\pi}\mathrm{Tr} \, \mathrm{Im} \, \mathbf{G}^{\mathrm{r}}(\omega)
\end{equation}
where the retarded Green's function is evaluated using the single-particle effective Hamiltonian, $\mathbf{h}_{\mathrm{eff}}\equiv\mathbf{h}-\frac{i}{2}\mathbf{\Gamma}$,
\begin{equation}\label{eq:G^r}
\mathbf{G}^{\mathrm{r}}(\omega)=(\omega\mathbf{I}-\mathbf{h}_{\mathrm{eff}})^{-1}.
\end{equation}
The spectral function is shown in Fig.~\ref{fig:spectral} for the four cases considered in Fig. \ref{fig:correlations} of the main text.

\section{Long-time limit}\label{sec:long_t}
In order to compute the current correlation function $C_{LR}(t_1,t_2)$ at stationarity for varying relative times $\tau$, we verify at which times the time-dependent currents saturate. As shown in Ref.~\cite{tuovinen2019distinguishing} with topological states, this is not as trivial as simply estimating the effective lifetime from $\sim\Gamma^{-1}$, but there can be long-lasting oscillations and saturation due to weak hybridization of the zero modes. The time-dependent net currents through the nanowire are shown in Fig.~\ref{fig:gamma0.01} for $\Gamma=\{0.01,0.001\}$. Based on these observations, we assign the long-time limits $t_1=t_2\equiv t=5000 \ J^{-1}$ for $\Gamma=0.01J$ and $t=15000 \ J^{-1}$ for $\Gamma=0.001J$. These are used for the calculation of the current correlation functions throughout.

\begin{figure}[h!]
  \includegraphics[width=0.6\linewidth]{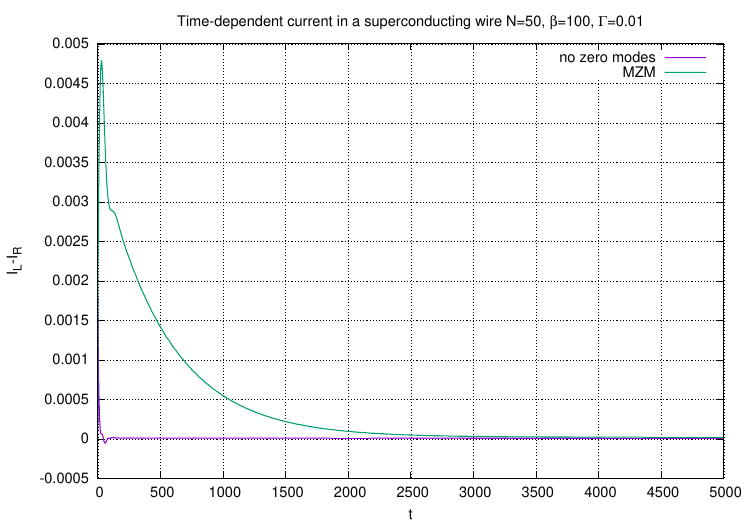}
~
  \includegraphics[width=0.6\linewidth]{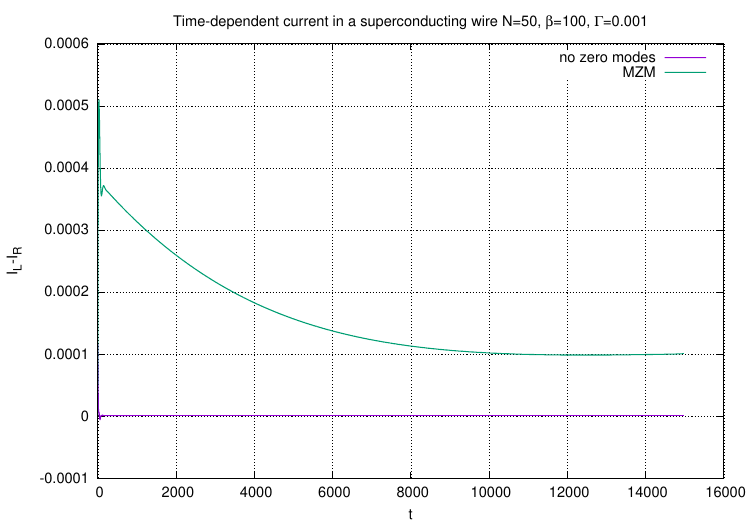}
  \caption{Time-dependent net currents through $N=50$ length nanowires at the ordinary and topological superconducting phases. The upper panel has the tunneling rate $\Gamma=0.01J$ while the lower panel has $\Gamma=0.001J$. The temperature is set by $\beta=100 \ J^{-1}$ for both cases.}
  \label{fig:gamma0.01}
\end{figure}

\section{Dependence on temperature and the tunneling rate of the leads}\label{sec:temperature_dep}
In addition to Fig.~\ref{fig:correlations} in the main text, here, we investigate the current cross-correlations' dependence on temperature and tunneling rate. The corresponding plots are shown in Fig.~\ref{fig:beta20}. We observe that there is little difference in the overall structure of periodically repeating peaks, i.e., we can conclude that the traversal-time characteristic is not highly sensitive toward temperature or the coupling strength.

\begin{figure}[t!]
  \includegraphics[width=0.45\linewidth]{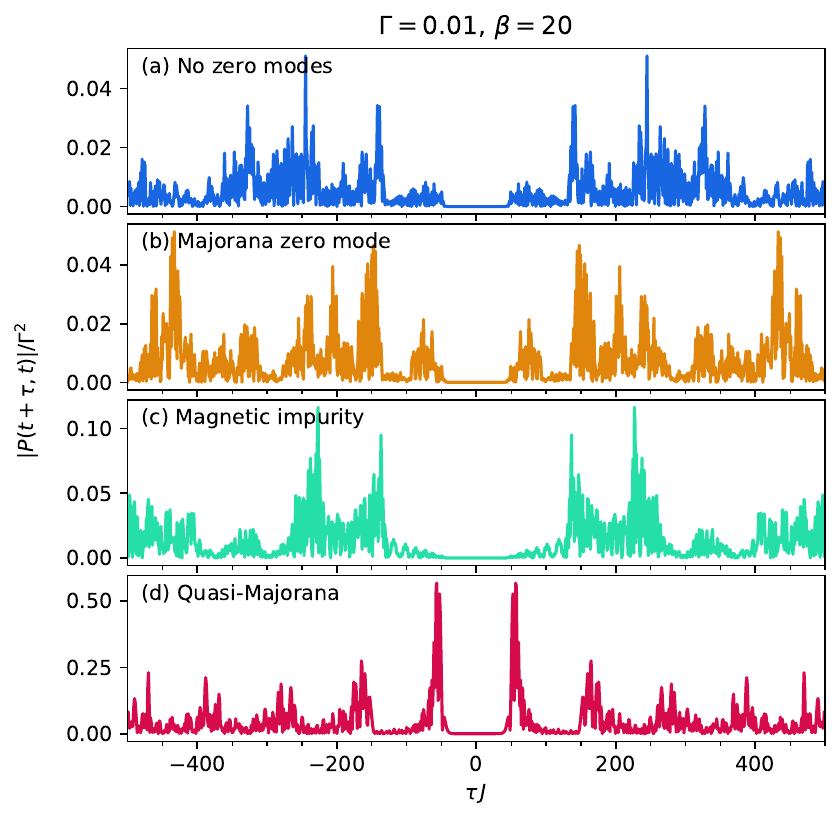}
  ~
  \includegraphics[width=0.45\linewidth]{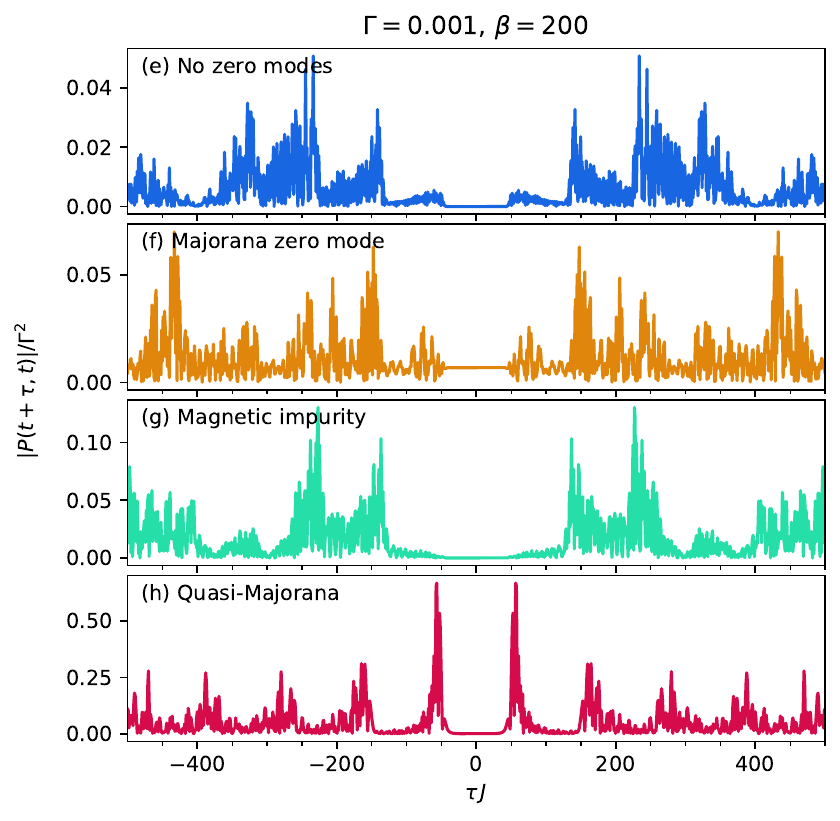}
  \caption{Current cross correlations for the four cases studied in the main text with otherwise identical parameter sets (cf.~Fig.~\ref{fig:correlations} in the main text) but with $\beta=20 \ J^{-1}$ in panels (a-d) and $\Gamma=0.001 J$ in panels (e-h).}
  \label{fig:beta20}
\end{figure}

\section{Identifying current cross correlation peaks and the traversal time: length-dependence}\label{sec:length}
As demonstrated in the main text, the traversal-time characteristic linearly depends on the nanowire length. Here, we provide the corresponding current cross correlation plots for nanowires of varying length. Fig.~\ref{fig:sites-ordinary} shows the cases of ordinary and topological superconductors, and Fig.~\ref{fig:sites-impurity} shows the cases of magnetic impurity and the quasi-Majorana state. The distance between the current cross correlation peaks clearly increases with the nanowire length.

\begin{figure}[h!]
  \includegraphics[width=0.4\linewidth]{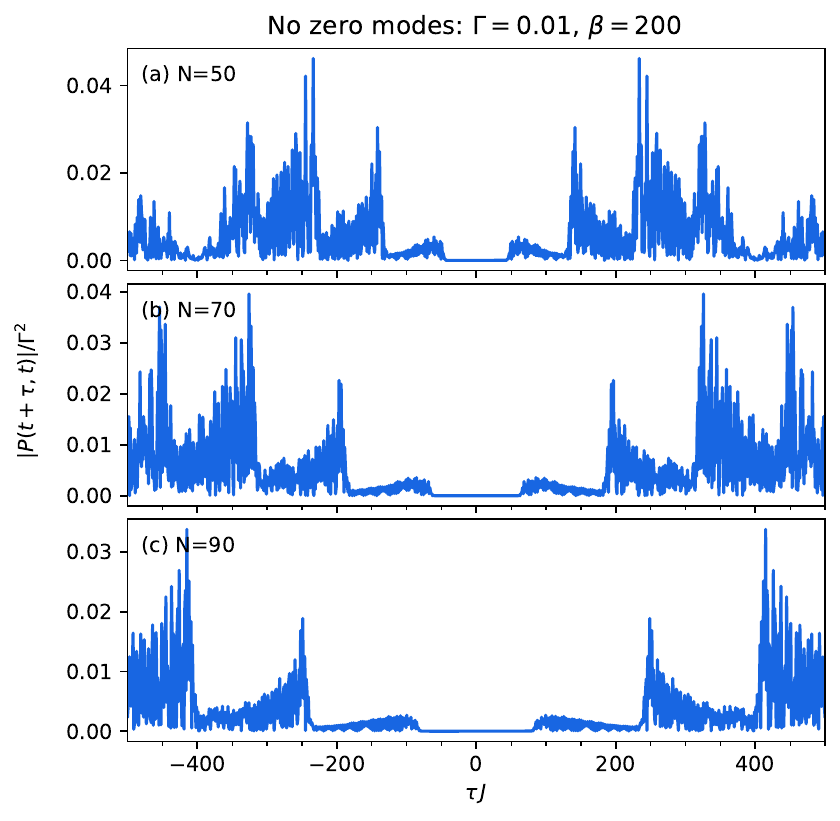}
  ~
  \includegraphics[width=0.4\linewidth]{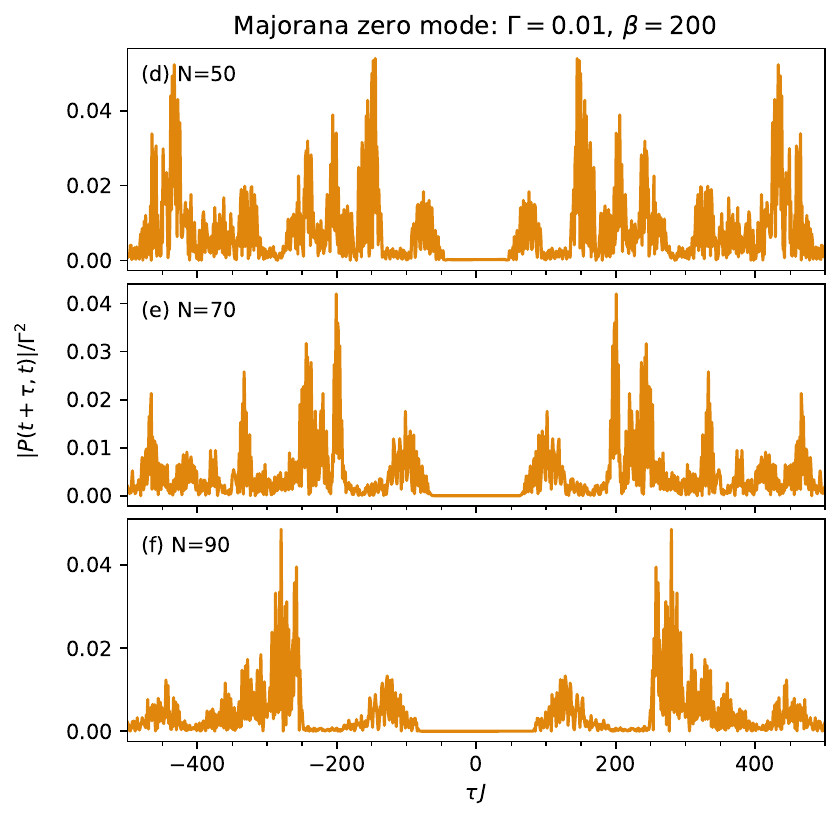}
  \caption{Current cross correlations for varying-length nanowires. Panels (a-c) show the ordinary superconductor while panels (d-f) show the topological superconductor.}
  \label{fig:sites-ordinary}
\end{figure}

\begin{figure}[h!]
  \includegraphics[width=0.4\linewidth]{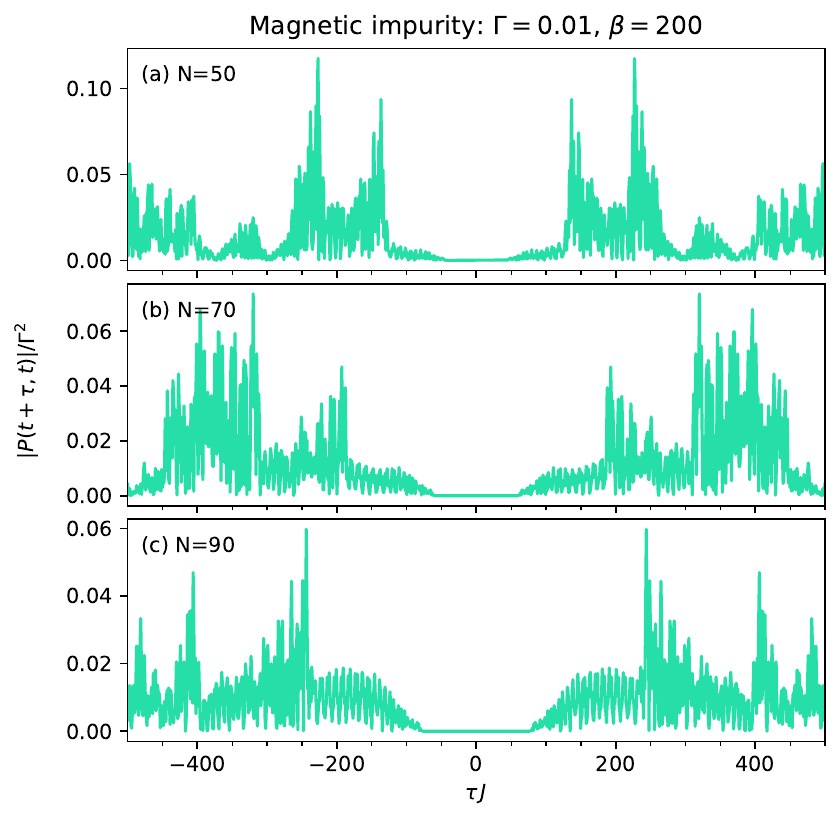}
  ~
  \includegraphics[width=0.4\linewidth]{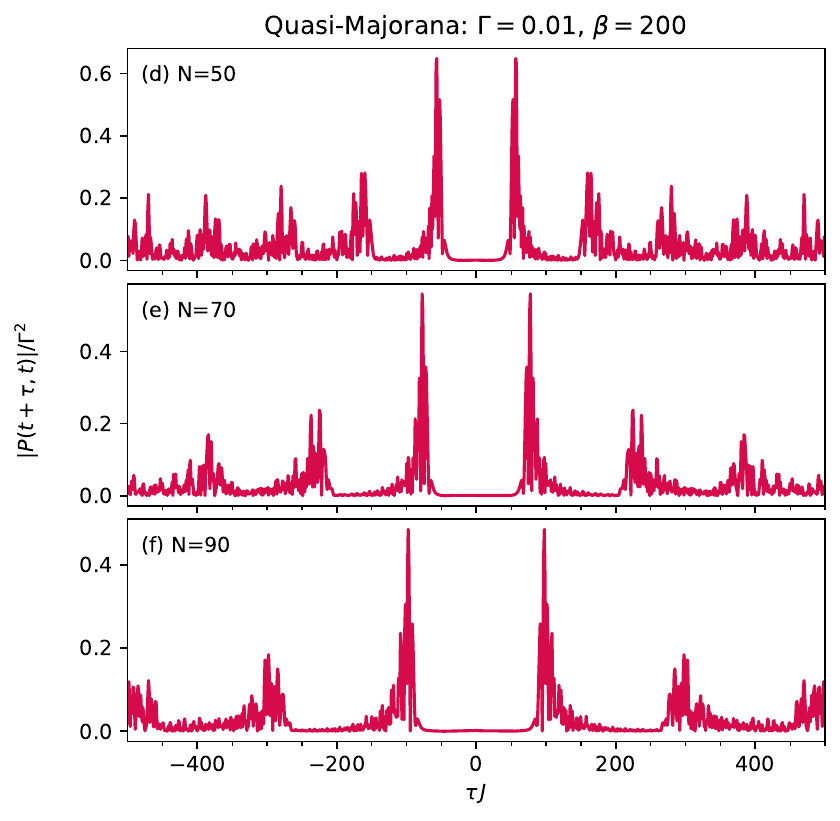}
  \caption{Same as Fig.~\ref{fig:sites-ordinary} but panels (a-c) show the magnetic impurity whereas panels (d-f) show the quasi-Majorana state.}
  \label{fig:sites-impurity}
\end{figure}

Now that we have seen above how the current cross correlation peak structure depends on the nanowire length, we can automatize identifying the corresponding peaks. This procedure is outlined as follows. First, by applying the Savitzky-Golay filter~\cite{savitzky_smoothing_1964} to the raw cross correlation data the signal is smoothed. Second, by applying Lorentzian fits to the first peaks around $\tau=0$ we may identify the central value of the Lorentzian as the traversal time and the half-width-half-maximum as the statistical error. The distance between the first main peaks, defined in the main text as occurring in the restricted domain $\mathcal{T}=[N/J,2N/J]$, is therefore $2\tau_{\mathrm{tr}}$. The procedure is visualized in Fig.~\ref{fig:ordinary-peak}. Extracting the associated traversal times and plotting them against the nanowire length is presented in Fig.~\ref{fig:traversal-length} of the main text where we observe a clear linear dependence. 

\begin{figure}[h!]
  \includegraphics[width=0.3\linewidth]{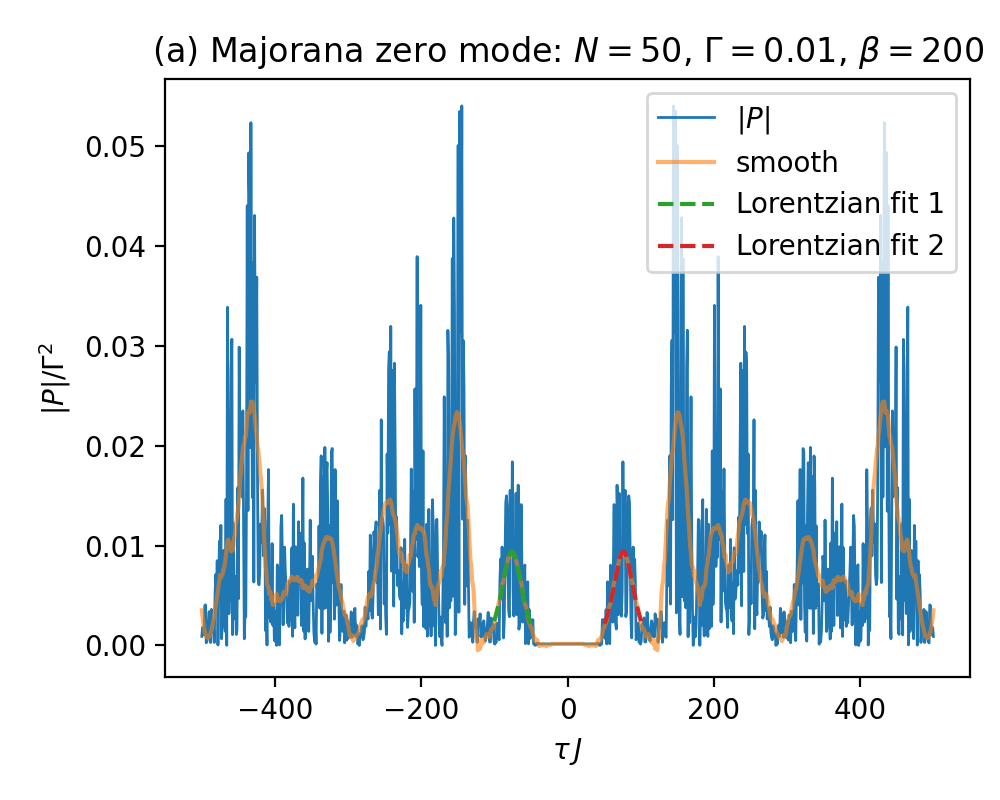}
  ~
  \includegraphics[width=0.3\linewidth]{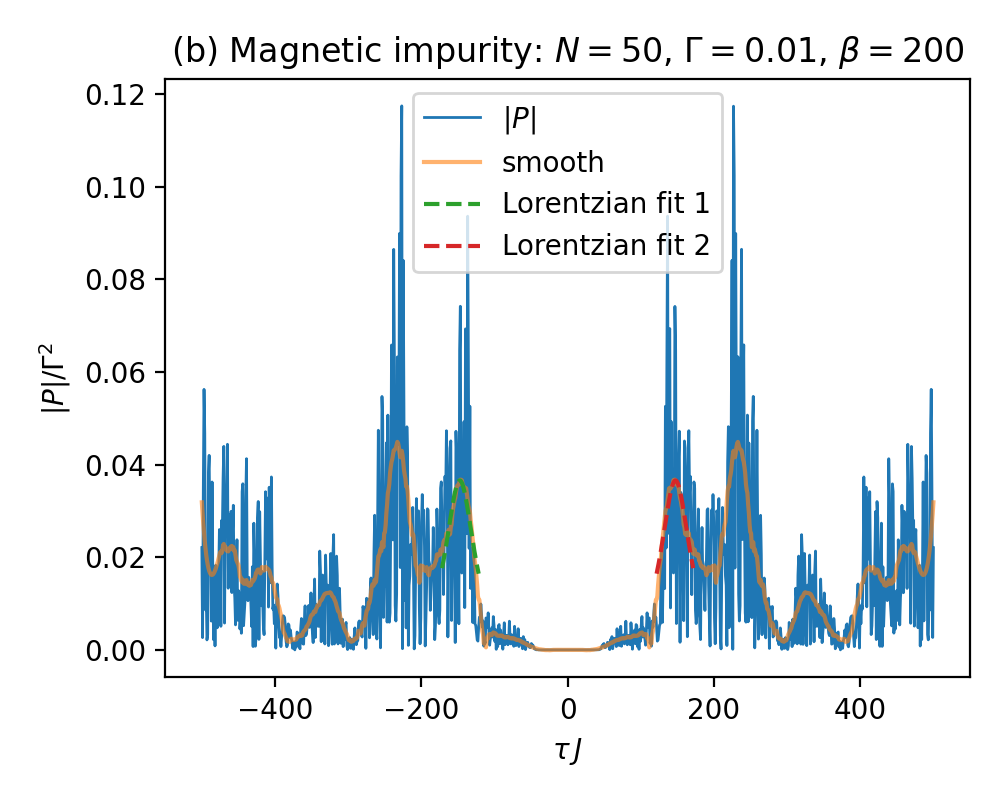}
  ~
  \includegraphics[width=0.3\linewidth]{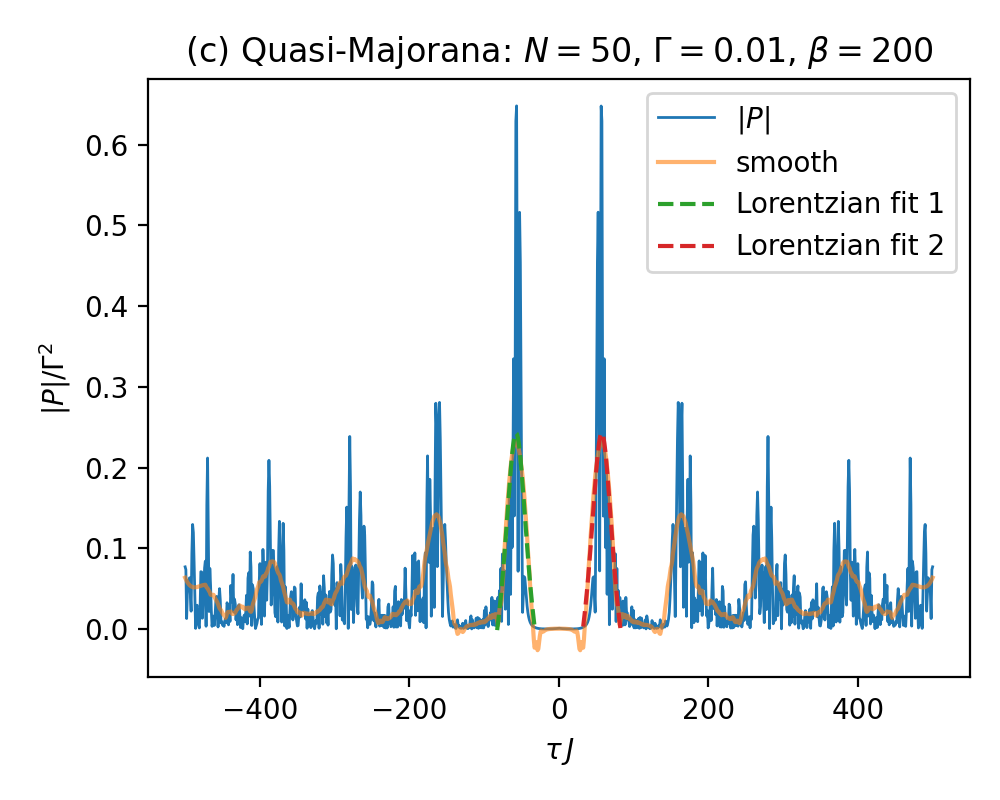}
  \caption{Identifying the current cross correlation peaks and the associated traversal times for the Majorana zero mode (a), magnetic impurity (b), and quasi-Majorana (c). The data correspond to $N=50$ length nanowires.}
  \label{fig:ordinary-peak}
\end{figure}

\section{Heuristic model for the traversal time}\label{sec:Appendix_heuristic}
\subsection{Site-by-site traversal time linear in $N$}
We wish to evaluate the traversal time $\tau_{\mathrm{tr}}$ as a site-by-site summation over individual site-local dwell times $\tau_{i}$:
\begin{equation}
\tau_{\mathrm{tr}}=\sum^{N}_{i=1}\tau_{i}.
\end{equation}
We do this by setting $\tau_{i}=1/v_{i}$, where $v_{i}$ is the site-local effective electron velocity in the superconducting wire. In general, this velocity should be proportional to the intramolecular coupling $\lambda_{i}$, since for strong coupling the rate of charge transfer through the wire is increased. Due to the end-site localization of the Majorana zero modes, with an exponential localization length drop-off of the on-site potential equal to $\zeta$, we can construct a piecewise approximation to this:
\begin{equation}\label{eq:lambda_i}
\lambda_{i}=\frac{J}{2}\left[\begin{array}{cc}
\exp\left[-\delta\left(\frac{z-i}{z}\right)\right] & 1\leq i\leq z\\
1 & z+1 \leq i \leq N-z\\
\exp\left[\delta\left(\frac{N-z+1-i}{z}\right)\right] & N-z+1\leq i\leq N
\end{array}\right.,
\end{equation}
where $z$ is the nearest integer approximation to $\zeta$, $\frac{J}{2}$ is the nearest-neighbour hopping in the absence of localization and $\delta$ controls the drop-off in localization in the end zones. Note that the absolute values of the end-site contributions (the $i=1$ and $i=N$ terms) are both equal to $\frac{J}{2}\exp(-\delta (\frac{z-1}{z}))$ in this construction, and the $i=z$ and $i=N-z+1$ terms are both $\frac{J}{2}$, guaranteeing continuity in the value of $\lambda_{i}$ across the nanowire. The exponential localization of the zero-energy states is shown in Fig.~\ref{fig:localization}, where the localization length $\zeta$ is extracted by an exponential fit $\sim\exp(-L/\zeta)$.

\begin{figure}[h!]
  \includegraphics[width=0.45\linewidth]{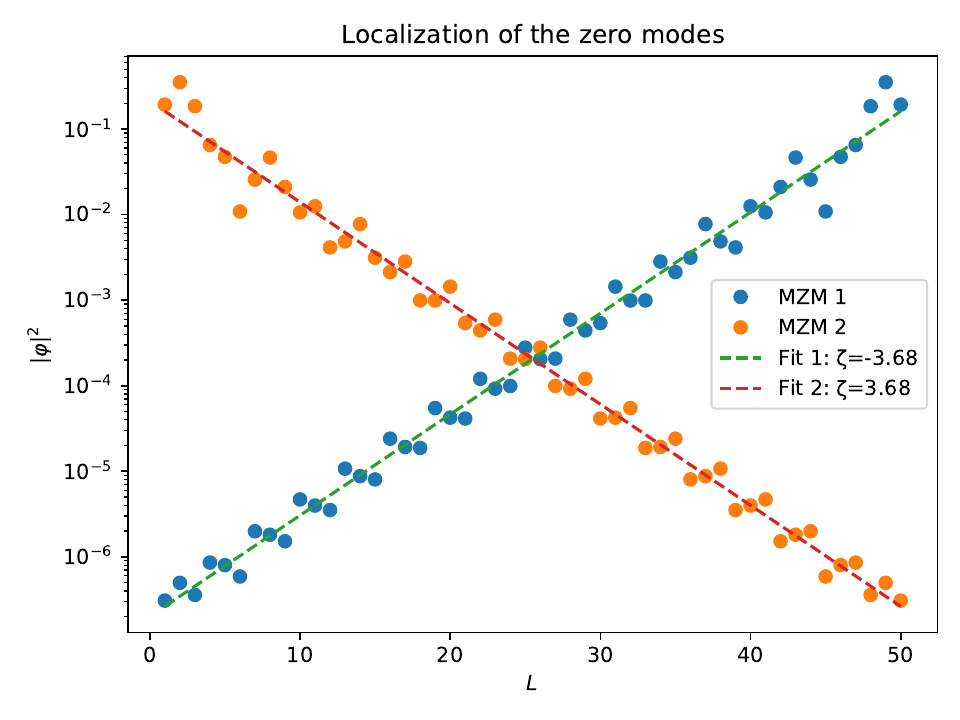}
  \caption{The probability density of the Majorana zero mode (two states pinned to zero energy) shows exponential localization around the nanowire edges. The localization length $\zeta$ is obtained by an exponential fit.}
  \label{fig:localization}
\end{figure}

Then, letting $v_{i}=\kappa\lambda_{i}=\frac{\kappa J}{2} s_{i}$, where $s_{i}$ is the site-dependent part of Eq. \eqref{eq:lambda_i}, we have for the full traversal time:
\begin{equation}\label{eq:ttr-expand}
\tau_{\mathrm{tr}}=\frac{2}{J\kappa}\left[\sum^{z}_{i=1}\exp\left[\delta\left(\frac{z-i}{z}\right)\right]+N-2z+\sum^{N}_{i=N-z+1}\exp\left[-\delta\left(\frac{N-z+1-i}{z}\right)\right]\right].
\end{equation}
Note that this expression is linear in the wire length $N$. 

The two summations in Eq.~\eqref{eq:ttr-expand} can be put into closed form using basic geometric series formulae, giving the following exact result:
\begin{equation} 
\tau_{\mathrm{tr}}=\frac{2}{J\kappa}\left[\frac{e^{\delta}-1}{e^{\delta/z}-1}+N-2z+\frac{e^{\delta}-1}{e^{\delta/z}-1}\right],
\end{equation}
which simplifies further to
\begin{equation}
    \tau_{\mathrm{tr}}=\frac{2}{J\kappa}\left[2S_{z}\left(\delta\right)+N-2z\right],
\end{equation}
where the end-site contribution is 
\begin{equation}
    S_{z}\left(\delta\right) = \sum^{z}_{i=1}\exp\left[\delta\left(\frac{z-i}{z}\right)\right] = \frac{e^{\delta}-1}{e^{\delta/z}-1} = \sum^{N}_{i=N-z+1}\exp\left[-\delta\left(\frac{N-z+1-i}{z}\right)\right].
\end{equation}

At this point, there are two free parameters, $\delta$ and $\kappa$, for the traversal time heuristic. Next, we will argue how $\kappa$ can be estimated from transport characteristics in the one-dimensional chain.

\subsection{Estimate of the gradient from the average electron velocity}
We now look to estimate a numerical value for the scaling factor $\kappa$, which determines the gradient of $\tau_{\textrm{tr}}$ with increasing wire length. To this end, we note that the Landauer-B{\"u}ttiker current in the steady-state is given by~\cite{buttiker_generalized_1985, ridley2022many}
\begin{equation}\label{eq:LB}
    I=\frac{e}{\pi}\overset{\infty}{\underset{-\infty}{\int}}d\omega\left[f\left(\omega,\mu_{L}\right)-f\left(\omega,\mu_{R}\right)\right]T\left(\omega\right) 
    = \frac{e}{\pi}\overset{\pi/a}{\underset{-\pi/a}{\int}}dk\left[f\left(\omega\left(k\right),\mu_{L}\right)-f\left(\omega\left(k\right),\mu_{R}\right)\right]T\left(k\right)\frac{d\omega\left(k\right)}{dk}.
\end{equation}
The first equality in Eq.~\eqref{eq:LB} is the standard energy integral containing (i) a difference in Fermi functions $f\left(\omega\left(k\right),\mu_{\gamma}\right)$, where $\mu_{\gamma}$ is the chemical potential of lead $\gamma$ and $\omega\left(k\right)$ is the dispersion relation, and (ii) the transmission function 
\begin{equation}
    T\left(\omega\right)=\textrm{Tr}\left[\mathbf{\Gamma}_{L}\mathbf{G}^{\mathrm{r}}\left(\omega\right)\mathbf{\Gamma}_{R}\mathbf{G}^{\mathrm{a}}\left(\omega\right)\right],
\end{equation}
where the retarded Green's function is defined in Eq. \eqref{eq:G^r} and the advanced Green's function is obtained as $\mathbf{G}^{\mathrm{a}}\left(\omega\right)=\left[\mathbf{G}^{\mathrm{r}}\left(\omega\right)\right]^{\dagger}$.

The second equality in Eq.~\eqref{eq:LB} relates the group velocity $v_{g}\left(k\right)=\frac{d\omega\left(k\right)}{dk}$ to the current. The current is thus evaluated as ``charge $\times$ population $\times$ velocity'', where $T(k)\equiv T\left(\omega\left( k\right)\right)$ measures the probability for a mode to be transmitted with momentum $k$. We recall that the electron charge $e=1$ in our unit system.

For a 1D chain of length $N$ with nearest neighbour hopping of $\frac{J}{2}$ and on-site energy $E_{0}$, the dispersion relation is given by
\begin{equation}\label{eq:disp}
    \omega\left( k\right)=E_{0}-J\textrm{cos}\left( ka\right).
\end{equation}
For terminal sites of the nanowire being coupled to wide-band leads, we have $\left[\mathbf{\Gamma}_{L}\right]_{ij}=\delta_{ij}\delta_{i1}\Gamma_{L}$ and $\left[\mathbf{\Gamma}_{R}\right]_{ij}=\delta_{ij}\delta_{iN}\Gamma_{R}$, so that the transmission is given by
\begin{equation}
    T\left(\omega\right)=\Gamma_{L}\Gamma_{R}|\left[\textbf{G}^{\textrm{r}}\left(\omega\right)\right]_{1N}|^{2}.
\end{equation}

To obtain an explicit functional form for $[\textbf{G}^{\textrm{r}}\left(\omega\right)]_{1N}$, one is required to invert the $N \cross N$ matrix (note that here we are not considering the Nambu basis)
\begin{equation}
\textbf{D}_{N} = \left(\omega\mathbf{I}-\mathbf{h}-i\frac{\mathbf{\Gamma}_{L}}{2}-i\frac{\mathbf{\Gamma}_{R}}{2}\right) \\
= \begin{pmatrix}
\omega-E_{0}-i\frac{\Gamma_{L}}{2} & -\frac{J}{2} & 0 & \ldots & 0\\
-\frac{J}{2} & \omega-E_{0} & -\frac{J}{2} & \ldots & 0\\
0 & -\frac{J}{2} & \omega-E_{0} & \ldots & 0\\
\vdots & \vdots & \vdots & \ddots & \vdots\\
0 & 0 & 0 & 0 & \omega-E_{0}-i\frac{\Gamma_{R}}{2}
\end{pmatrix},
\end{equation}
where $\mathbf{h}$ is the Hamiltonian matrix block corresponding to the `bare' nanowire. This can be done using Cramer's rule, leading to the expression
\begin{equation}
    [\textbf{G}^{\textrm{r}}\left(\omega\right)]_{1N}=\frac{(\frac{J}{2})^{N-1}}{|\textbf{D}_{N}|},
\end{equation}
where $|\textbf{D}_{N}|$ is the determinant of $\textbf{D}_{N}$, and we note that the $1N$-component of $\textbf{D}_{N}$ has $-\frac{J}{2}$ along the main diagonal. 

Expanding the determinant using the dispersion relation~\eqref{eq:disp}, the transmission function can be written as
\begin{equation}
    T\left(k\right)=\frac{\Gamma_{L}\Gamma_{R}}{J^{2}}\frac{\sin^{2}\left(k\right)}{\left(\sin^{2}\left(k\right)+\cos^{2}\left(k\right)\sin^{2}\left(Nk\right)\right)}.
\end{equation}
The $k$-averaged electron velocity is then given by
\begin{gather}\label{eq:v_kintegral}
    v_{\mathrm{av}}	= \frac{\kappa J}{2} = \frac{\overset{\pi}{\underset{0}{\int}}dk \, T\left(k\right)v_{g}\left(k\right)}{\overset{\pi}{\underset{0}{\int}}dk \, T\left(k\right)}
	=\frac{J\overset{\pi}{\underset{0}{\int}}dk \, \frac{\sin^{3}\left(k\right)}{\left(\sin^{2}\left(k\right)+\cos^{2}\left(k\right)\sin^{2}\left(Nk\right)\right)}}{\overset{\pi}{\underset{0}{\int}}dk \, \frac{\sin^{2}\left(k\right)}{\left(\sin^{2}\left(k\right)+\cos^{2}\left(k\right)\sin^{2}\left(Nk\right)\right)}}
    \equiv J\frac{K_{1}}{K_{2}}.
\end{gather}

\begin{table}[htbp]
\centering
\begin{tabular}{ |p{1.0cm}||p{1.0cm}|p{1.0cm}|p{1.0cm}|  }
 \hline
 \multicolumn{4}{|c|}{Numerical Integrals} \\
 \hline
 $N$ & $K_{1}$ & $K_{2}$ & $K_{1}/K_{2}$ \\
 \hline
 10   & 1.5704    & 1.9840 & 0.7916 \\
 20 &  1.5708  & 1.9959   & 0.7870  \\
 30 & 1.5708 & 1.9981 & 0.7861\\
 40    & 1.5708 & 1.9990 & 0.7858\\
 50 & 1.5708 & 1.9993 & 0.7857\\
 60    & 1.5708 & 1.9995 & 0.7856\\
 70 & 1.5708 & 1.9997 & 0.7855\\
 80    & 1.5708 & 1.9997 & 0.7855\\
 90 & 1.5708 & 1.9998 & 0.7855\\
 100    & 1.5708 & 1.9998 & 0.7855\\
 200    & 1.5708 & 2.0000 & 0.7854\\
 \hline 
\end{tabular}
\caption{Convergence of the numerical integrals in Eq.~\eqref{eq:v_kintegral} for $N$ lattice points in the nanowire.}
\label{tab:k_integrals}
\end{table}

The $k$-integrals $K_{1}$ and $K_{2}$ can be evaluated numerically for different $N$. Table~\ref{tab:k_integrals} displays numerical values for $K_{1}$, $K_{2}$ and their ratio for $N$ ranging from $10$ to $200$, demonstrating a convergence to $0.7854\approx\pi/4$ in the large $N$ limit. Finally, we see that, for $N\geq50$, $v_{\mathrm{av}} \approx \frac{J\pi}{4}$, i.e. $\kappa \approx \frac{\pi}{2}$, and we can fit the traversal time to 
\begin{equation}
    \tau_{\mathrm{tr}}=\frac{4}{J\pi}\left[2S_{z}\left(\delta\right)+N-2z\right].
\end{equation}

\end{widetext}

\end{document}